\documentstyle[12pt]{ioplppt}
\input psfig.sty
\def\pin{$\mbox{}$\indent}  %force to indent the first paragraph

\def\bsigma{\mbox{\boldmath$\sigma$}}
\def\erf{\mbox{erf}}
\def\RS{(RS)}
\def\RSB1{(RSB1)}

\begin{document}
\title{A replica approach to the state-dependent synapses neural network}
\author{
D.~Boll\'e\ftnote{3}{E-mail: desire.bolle@fys.kuleuven.ac.be}, 
G.M.~Shim\ftnote{1}{E-mail: gmshim@nsphys.chungnam.ac.kr} and
B.~Van Mol\ftnote{2}{E-mail: bart.vanmol@fys.kuleuven.ac.be}}
\address{\S\ddag\ Instituut voor Theoretische Fysica,
            K.U.\ Leuven, B-3001 Leuven, Belgium }
\address{\dag\
            Department of Physics, Chungnam National University, 
            Yuseong, Taejon 305-764, R.O.Korea}
\begin{abstract}
\noindent
The replica method is applied to a neural network model with 
state-dependent synapses built from those patterns having a correlation 
with the state of the system greater than a certain threshold. 
Replica-symmetric and first-step replica-symmetry-breaking results are 
presented for the storage capacity at zero temperature as a function of
this threshold value. 
A comparison is made with existing results based upon mean-field 
equations obtained by using a statistical method.
\end{abstract}
\date{}

%PACS numbers: 75.10.Nr, 05.20.-y, 64.60.Cn

\section{INTRODUCTION}
\pin

It is standard knowledge by now that for the Hopfield model \cite{Ho}
 the Hebb rule
leads to a critical storage capacity $\alpha_c= 0.138$ \cite{AGS} while 
for these type of models with quadratic interaction the optimal storage 
capacity is $\alpha_c= 2$ \cite{Ga}. This is
due to the fact that the contribution of the noise caused by the weakly 
correlated patterns becomes larger than the signal of the condensed
patterns as $\alpha$ increases. In order to lift this limitation of the 
Hebb rule a model 
with state-dependent synapses has been discussed recently 
\cite{MP,ZLW1}. The idea thereby is to introduce a threshold $\eta$
cutting out of the Hebb rule all patterns whose correlations with the
state of the system are smaller than this threshold \cite{ZLW1}.
These authors propose an energy function for this state-dependent
synapse (SDS) model and derive the corresponding
fixed-point equations using the so-called heuristically motivated 
statistical mean-field scheme developed
in \cite{Ge,Pe} (see also \cite{HKP}). In the case of the Hopfield model
($\eta=0$) this statistical derivation leads to the same results as
those derived  using a replica symmetric mean-field theory
approach \cite{AGS}. Solving these fixed-point equations one 
finds, for example at zero temperature,
an increase in the storage capacity from  $\alpha_c= 0.138$ for $\eta=0$
up to, e.g., $\alpha_c= 0.17$ for $\eta=1$. A similar effect has been found 
for the recognition of temporal sequences \cite{ZLW2} and for
non-monotonic Hopfield models \cite{In}.

In this paper we apply the replica method to the zero temperature
capacity problem of the SDS model. The aim thereby is twofold.
First we want to find out whether the replica symmetric (RS) fixed-point
equations derived by the standard replica approach again coincide with
the fixed-point equations found with the statistical method. Second
since we expect that the RS results are unstable at zero temperature, we
want to determine the effects of a first-step replica-symmetry breaking
(RSB1) on the capacity.    

Somewhat surprisingly, we find that the RS fixed-point equations are
different
from the results obtained in \cite{ZLW1}. This is due to the fact that
for $\eta \neq 0$ the assumptions made in \cite{ZLW1} that both the 
overlap 
with the non-condensed patterns as well as the noise induced by these
non-condensed patterns have a Gaussian distribution are incompatible.
Keeping only the (standard) assumption that the noise is Gaussian we 
can improve the calculations using the statistical scheme and show 
agreement with the replica symmetric approach.
Furthermore, in an RSB1 treatment the critical storage capacity
increases versus the RS values but up to $\eta=1$ the increase is
relatively small. 

The rest of this paper is organized as follows. In section 2 the SDS-model
is shortly reviewed. In section 3 the RS approach to this model at zero
temperature is outlined and a detailed comparison with the statistical 
method used in \cite{ZLW1} is made. Section 4 contains a discussion of 
the RSB1 solution. Some concluding remarks are given in section 5.

\section{The SDS-model}
\pin
Consider a network of $N$ neurons which can take the values $\pm 1$ with
equal probability. In this network we want to store $p=\alpha N$ patterns 
$\xi_i^\mu = \pm 1,\, i=1,2,\ldots,N, \,\mu =1,2,\ldots,p$ that 
are supposed to be independent and identically distributed random 
variables with probability distribution 
$\mbox{Pr}(\xi_i^\mu)=
\frac{1}{2}\delta(\xi_i^\mu -1)+\frac{1}{2}\delta(\xi_i^\mu + 1)$.
                                        
Given a configuration $\bsigma = (\sigma_1,\ldots,\sigma_N)$, the
local field $h_i$ of neuron $i$ is
\begin{equation}
    h_i(\bsigma) = \sum_{j \neq i} J_{ij}\sigma_{j} \,,
    \label{eq:field} 
\end{equation}
where $J_{ij}$ are the synaptic couplings given by
\begin{equation}
   J_{ij}= \frac{1}{N} \sum_{\mu =1}^{\alpha N} \xi_i^\mu\xi_j^\mu
      \Theta \left((m^\mu)^2-\frac{\eta^2}{N} \right)
     \label{eq:couplings} 
\end{equation}
with $m^\mu$ the usual overlap order parameters defined by
\begin{equation}
    m^\mu \equiv \frac{1}{N} \sum_{i=1}^N\xi_i^\mu \sigma_i
     \label{eq:overlap}
\end{equation}
and $\eta \geq 0$ the threshold parameter. Due to the presence of the   
step function $\Theta(\cdot)$ only those terms where 
$(m^\mu)^2\geq \eta^2/N$ contribute to the synaptic couplings.  

The neurons are updated asynchronously according to the well-known
Glauber dynamics. We are interested in the zero temperature limit of
this dynamics, which can be written as 
\begin{equation}
   \sigma_i(t+1) = \mbox{sgn} [h_i(\bsigma(t))] \,.
    \label{eq:dyn}
\end{equation} 
For this deterministic dynamics an energy function has been found in 
ref.~\cite{ZLW1} 
\begin{equation}
   H = -\frac{N}{2}\sum_{\mu} \left((m^\mu)^2-\frac{\eta^2}{N} \right)
            \Theta\left((m^\mu)^2-\frac{\eta^2}{N} \right) \,.
	\label{eq:lya}  
\end{equation}

\section{A replica approach}
\pin
The energy function (\ref{eq:lya}) has been used in \cite{ZLW1} 
to derive fixed-point equations for the relevant order parameters using
the statistical mean-field scheme \cite{Ge,Pe}. The key idea of the
latter calculation is to assume that the noise to which the small overlaps
with the non-condensed patterns add up is Gaussian.

In the following we apply the replica approach \cite{AGS,HKP} at
zero temperature up to first-order breaking and compare our results with 
the statistical scheme.

\subsection{Replica symmetric results}
\pin
Following the standard approach we calculate the replica-symmetric 
free energy per neuron for the SDS-model at zero temperature as
 the limit $\beta \rightarrow \infty$, with $\beta$ the
inverse temperature, of its temperature dependent form. 
For the latter we obtain as a function of the usual order parameters,     
i.e.,  the overlap, $m^1=m$, with the condensed pattern $\mu=1$, the
Edwards-Anderson order parameter, $q$, and the residual overlap, $r$, with
the non-condensed patterns $\mu \geq 2$,  
\begin{equation}
    f^{\RS}(m,r,q,\beta)= f_{0}^{\RS}(m,r,q,\beta) 
                 + f_{\eta}^{\RS}(q,\beta) \,,
    \label{eq:fe}
\end{equation} 
where
\begin{eqnarray} 
    && \hspace{-0.9cm}
    f_{0}^{\RS}(m,r,q,\beta) = \frac{m^2}{2}+\frac{1}{2}\alpha \beta r(1-q)
        + \frac{\alpha}{2 \beta} 
	  \left[ \ln(1-\beta(1-q)) - \frac{\beta q}{1-\beta(1-q)}
	                 \right]  \nonumber \\
     && \hspace{-0.9cm}
     -\frac{1}{\beta} 
          \int\, Dz \ln [2 \,\mbox{cosh}\, \beta(m+\sqrt{\alpha  r}\,z)] 
	       \label{eq:fehop} \\
   && \hspace{-0.9cm}
     f_{\eta}^{\RS} (q,\beta)= \frac{\alpha}{2}\eta^2
         -\frac{\alpha}{\beta}
	  \int\, Dz \ln \left[1-\frac12 \erf \,(\phi_+(\beta))
	            -\frac12 \erf \,(\phi_-(\beta)) \right.
		    \nonumber \\
   &&\hspace{-0.9cm}
      \left. +\frac12 \sqrt{1-\beta(1-q)} \, 
	     \exp \left(\frac{\beta}{2}
	         (\eta^2-\frac{q z^2}{1-\beta(1-q)}) \right) 
	\left( \erf \, (\phi_+(0)) +\erf \, (\phi_-(0)) \right)\right]
	     \label{eq:fesds}
\end{eqnarray} 
with
\begin{equation}
   \phi_{\pm}(x) = \frac{[1- x(1-q)]\eta \pm \sqrt{q}\, z}
                        {\sqrt{2(1-q)(1- x(1-q))}} 
	\label{eq:phi}		
\end{equation}
and $Dz= dz (2\pi)^{(-1/2)} \exp(-z^2/2)$. In the above
$f_{0}^{\RS}(\cdot)$ is the free energy corresponding to the Hopfield
model ($\eta=0$) while $f_{\eta}^{\RS}(\cdot)$ reflects the effect of
the removal of the non-condensed patterns having a small correlation
with the state of the system. 
Furthermore, the fixed-point equations are given by 
\begin{eqnarray}
   m &=& \int \, Dz \, \mbox{tanh} \,\beta (m+\sqrt{\alpha r}\,z) 
           \label{eq:fixedm}     \\
   q &=& \int \, Dz \, \mbox{tanh}^2 \,\beta (m+\sqrt{\alpha r}\,z)
         \label{eq:fixedq}  \\
   r &=& \frac{q}{[1- \beta(1-q)]^2} 
        + \frac{2}{\alpha \beta} \frac{\partial} {{\partial q}}
	             f_{\eta}^{\RS}(q, \beta)
	\label{eq:fixedr}
\end{eqnarray}
In the limit $\eta \rightarrow 0$ we find back the
fixed-point equations for the Hopfield model, as we should. Furthermore,
the change in the Hebb rule realized in eq.~(\ref{eq:couplings})
manifests itself explicitly only in the order parameter $r$ as one
would expect.
For zero temperature 
the fixed-point equations (\ref{eq:fixedm})--(\ref{eq:fixedr}) reduce to
\begin{eqnarray} 
 m  &= & \mbox{erf}\left(\frac{m}{\sqrt{2 \alpha r}} \right) 
              \label{eq:zerom}  \\                 
 r  &= & \frac{1}{{(1-c)}^2}\left[1-\mbox{erf}
     \left(\sqrt{\frac{1-c}{2}}\eta \right)+
                   \sqrt{\frac{2(1-c)}{\pi}}\eta
                 \exp \left(-\frac{1}{2}(1-c)\eta^2 \right)\right]
	\label{eq:zeror} \\
 c & = &\lim_{\beta \rightarrow \infty} \beta(1-q)
    =  \sqrt{\frac{2}{\pi \alpha r}} 
               \exp{\left(-\frac{m^2}{2 \alpha r}\right)} \,.
	       \label{eq:zeroc}	
\end{eqnarray} 
These results have to be compared with the statistical mean-field scheme
of \cite{ZLW1}. 
But first we check the RS stability of this solution
(\ref{eq:zerom})-(\ref{eq:zeror}) by calculating, as an indication, the
 entropy of the replica symmetric phase. We find 
\begin{equation}
  S= S_{0} + S_{\eta}
  \label{eq:en}
\end{equation} 
where
\begin{eqnarray} 
   S_{0} &=& -\frac{\alpha}{2}\left[ \ln(1-c)+\frac{c}{1-c} \right] 
 \label{eq:enhop} \\ & & \nonumber \\ & & \nonumber \\
  S_{\eta} &=& \frac{\alpha}{2}\ln (1-c) \,
         \mbox{erf}\left(\sqrt{\frac{1-c}{2}}\eta \right)
	 \nonumber \\
     &&-\frac{\alpha  c}{{(1-c)}^2}
        \left[\sqrt{\frac{1-c}{2\pi}}\, \eta \,
	   \exp \left(-\frac{1}{2}(1-c)\eta^2 \right)
     -\frac{c}{2} \mbox{erf}\left(\sqrt{\frac{1-c}{2}}\eta 
                   \right) \right] 
    \label{eq:ensds}
\end{eqnarray}
Again in the limit $\eta \rightarrow 0$ the expression (\ref{eq:en})
reduces to the entropy of the Hopfield model (\ref{eq:enhop}) as
given, e.g, in \cite{AGS}. As shown in Fig.~1, the entropy for the
SDS-model is negative for all values of $\eta$ indicating RS-breaking.
We remark that for $\eta \rightarrow \infty$ the entropy goes to
$-\infty$ as $-{\eta}^{-1} \exp(\eta^2/2)$.
To get an idea about the size of the breaking as a function of $\eta$
a first-order approximation (RSB1) will be performed in section 3.3.
\begin{figure}[t]
\centerline{
\vspace{.2cm}
\psfig{figure=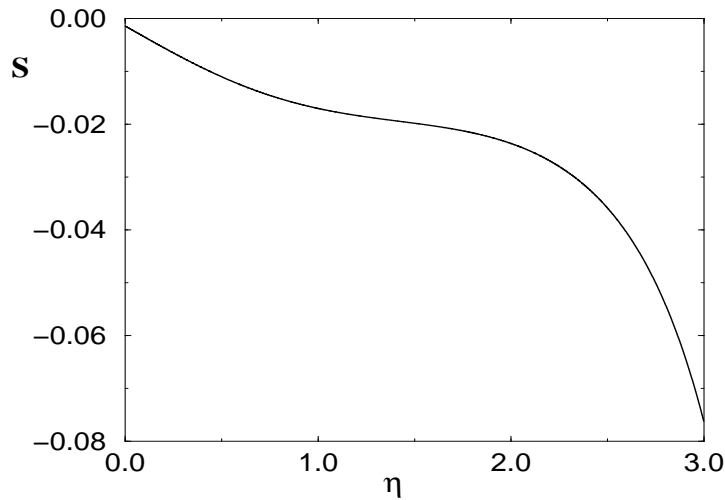,height=6.5cm,width=9.5cm,angle=270}}
\label{fig1}
\caption{The replica symmetric entropy $S$ at zero temperature as a
function of the threshold $\eta$.}
\end{figure}
\subsection{Comparison with the statistical scheme}
\pin
Comparing the fixed-point equations  (\ref{eq:zerom})-(\ref{eq:zeror})
with those of
the statistical mean-field scheme derived in \cite{ZLW1} (see
egs.~(23)-(26)) we find that they are different as soon as $\eta \neq 0$.
This is illustrated by Fig.~2.
It contrasts the situation for the Hopfield model where it is argued
\cite{Ge} that the key assumption in the statistical approach mentioned
above -- the noise to which the small overlaps with the non-condensed 
patterns add up is Gaussian -- oversimplifies and is most probably
responsible for obtaining the results corresponding to the RS
approximation. 
The reason that there is no such correspondence here is that the 
derivation in \cite{ZLW1}
not only invokes this key assumption but furthermore supposes that the
overlap with the non-condensed patterns themselves, i.e., the 
$\langle m^{\mu}_N \rangle, \mu=2,\ldots,p=\alpha N$ where the brackets
$ \langle \cdot \rangle$ indicate the thermal average, have an identical
normal distribution with mean
zero and variance $\sigma^2/N$. (For convenience we write down
explicitly the N-dependence in this subsection). These two assumptions 
are incompatible for $\eta \neq 0$. 
Indeed, following closely the derivation in 
\cite{ZLW1,Ge} by starting from the mean-field equations for
the thermal average of the overlap with a non-condensed
pattern, $\langle m^\nu_N \rangle$, and expanding it in a Taylor series
to first order we arrive at   
\begin{equation}
   \left[1-\beta (1-q_N) \Theta 
         \left({\langle m^\nu_N \rangle}^2 - \frac{\eta^2}{N} 
               \right)\right]
	       \langle m^\nu_N \rangle = X_N
	   \label{eq:verdeling}    
\end{equation} 
with 
\begin{eqnarray}
    X_N&=&\frac{1}{N}\sum_{j=1}^N\xi_j^\nu \xi_j^1\tanh
                         \beta(m_N+\eta_{N,j}^{\nu})
           \label{eq:xn}    \\
    q_N&=&\frac{1}{N}\sum_{j=1}^N{\tanh}^2 \beta(m_N+\eta_{N,j}^{\nu})
           \label{eq:qn}    \\   
   \eta_{N,j}^{\nu} &=& \sum_{\mu\neq 1,\nu}
        \xi_j^\mu \xi_j^1 \langle m^\mu_N \rangle 
            \Theta\left({\langle m^\mu_N \rangle}^2 - \frac{\eta^2}{N} 
                                        \right)  
    \label{eq:noise}
\end{eqnarray}
Here $\eta_{N,j}^{\nu}$ is the noise part and we recall that
 $\langle m^1_N \rangle \equiv  m_N $ is the overlap of the  
network  with the condensed pattern $1$.
\begin{figure}[t]
\centerline{
\vspace{.2cm}
\psfig{figure=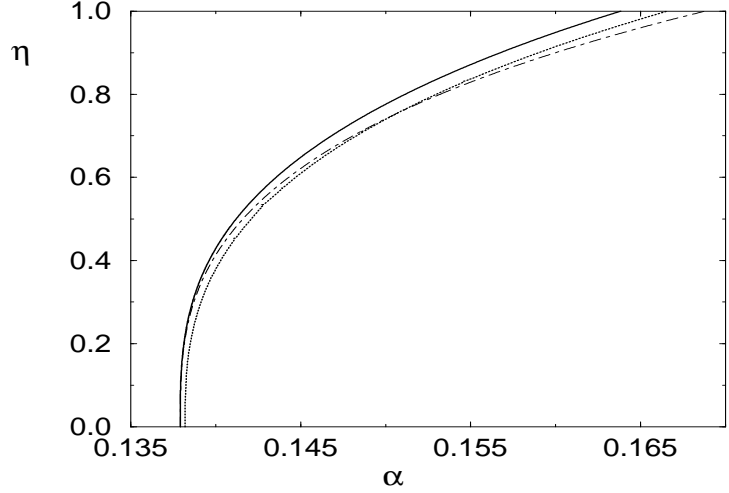,height=6.5cm,width=9.5cm,angle=270}}
\label{fig2}
\caption{The critical capacity in the RS (full curve) and RSB1 (dotted
curve) approximation as a function of the threshold $\eta $. For
comparison we also show the results of Ref.~[5] (dashed-dotted curve).}
\end{figure}
%{\epsfig{file=rsbrsmf.eps, height = 9cm,width=6cm, angle = 270}} \\ \\
This expression shows that in the limit $N \rightarrow \infty$ the
relation between the distributions for the random variables 
$\langle m^\nu \rangle$ and $\eta_{j}^{\nu}$ is no longer simply linear
 when $\eta \neq 0$.  
In fact, starting from the key assumption that the noise has a Gaussian
distribution, i.e., $\eta^{\nu} \sim {\cal{N}}(0,\alpha r)$ we find 
from (\ref{eq:xn}) using the central limit theorem that
 $\lim_{N \rightarrow \infty}\sqrt{N}X_N \sim {\cal{N}}(0,q)$ with
 $q$ the limit of (\ref{eq:qn}),
which is equal to eq.~(\ref{eq:fixedq}). Furthermore, according to
eq.~(\ref{eq:verdeling}) we see that $\langle m^\nu_N \rangle$ is a
multi-valued function of the $X_N$. Employing a standard geometrical 
Maxwell construction we obtain
\begin{equation}
  \sqrt{N}\langle m^\nu_N \rangle =
     \left\{ \begin{array}{ll}
          \frac{\sqrt{N}X_N}{1-\beta(1-q_N)}& \qquad  
     \left|\sqrt{N}X_N\right| > \sqrt{1-\beta(1-q_N)}\,\eta 
                   \\  & \\
            \sqrt{N} X_N & \qquad \left|\sqrt{N}X_N\right| 
                           < \sqrt{1-\beta(1-q_N)}\,\eta
        \end{array} \right.
	\label{eq:mdis}
\end{equation}
The expression (\ref{eq:mdis}) clearly shows that the overlaps with the
non-condensed patterns are not Gaussian distributed in the limit $N
\rightarrow \infty$. Using
the correct distribution we precisely find eq.~(\ref{eq:zeror}) for the
order parameter $r$ in the zero-temperature limit. Hence, we have shown 
complete equivalence between the statistical mean-field scheme using only
the key assumption that the noise is Gaussian distributed, and the
replica symmetric results.

\subsection{First-step breaking results}
\pin
From the observation on the entropy given in section 3.1 we expect RSB
effects. In order to get an idea about the size of these effects with
growing $\eta$ we apply first-step RSB.  We follow the standard approach
(see, e.g., \cite{MPV}) by introducing the order parameters
\begin{equation}
\begin{array}{ll}
    \begin{array}{lll}
       m^\mu_\alpha &=& m^\mu 
    \end{array} & \forall \alpha = 1,\cdots,n \\
    \begin{array}{lll}
    q_{\alpha \gamma} &=&(1-q_1)\delta_{\alpha \gamma} 
                  + (q_1-q_0)\epsilon_{\alpha \gamma} + q_0 \\
    r_{\alpha \gamma} &=&(1-r_1)\delta_{\alpha \gamma} 
                  +(r_1-r_0)\epsilon_{\alpha\gamma} + r_0 
    \end{array} &  \forall \alpha, \gamma = 1,\cdots,n 
\end{array} \label{eq:rsborder}
\end{equation}
with $n$ the number of replicas and  
$\{\epsilon_{\alpha \gamma}\},\forall \alpha, \gamma = 1,\cdots,n $ a 
$(n\times n)$ - matrix with elements 1 inside  $n/k$
diagonal blocks of size  $k$ and $0$  outside these blocks.

The free energy per neuron can then be obtained after some tedious 
calculations
\begin{eqnarray}
    \hspace*{-1.1cm}
    f^{\RSB1}(m,q_0,q_1,r_0,r_1,k,\beta) \nonumber \\
          = f^{\RSB1}_{0}(m,q_0,q_1,r_0,r_1,k,\beta) 
                    +f^{\RSB1}_{\eta} (q_0,q_1,k,\beta) \, ,
          \label{eq:frsb}		    
\end{eqnarray}
where the first term is given by 
\begin{eqnarray}
   &&\hspace*{-1.1cm}
   f^{\RSB1}_{0}(m,q_0,q_1,r_0,r_1,k,\beta)
     = \frac12 m^2 - \frac{\ln 2}{\beta}
           - \frac12\alpha \beta [k q_0r_0 +(1-k)q_1r_1 -r_1]
	   \nonumber \\
	&&+ \frac{\alpha}{2\beta}\ln[1-\beta(1-q_1)]  
           \nonumber \\   
	   &&- \frac{1}{k\beta}
	    \int Dz_1 \ln \left\{ \int Dz_2 \, \mbox{cosh}^k
       \left[\beta(m +\sqrt{\alpha r_0}\, z_1 
             +\sqrt{\alpha (r_1-r_0)}\, z_2)\right] \right\}
       \label{eq:f0rsb}
\end{eqnarray}  
and the second term reads    
\begin{eqnarray}    
  && \hspace*{-1.1cm}
  f^{\RSB1}_{\eta}(q_0,q_1,k,\beta) = \frac{\alpha}{2}\eta^2
       -\frac{\alpha}{\beta k} \int Dz_1\,\ln  \int Dz_2
            \nonumber \\
      && \left \{ 
       \exp\left[ \frac{\beta z^2}{2(1-\beta(1-q_1))} \right]  
                 [1-\frac12 \erf \,(\psi_+(\beta))
	            -\frac12 \erf \,(\psi_-(\beta)) ] \right.
		    \nonumber \\
         &&\hspace*{-1.1cm}
      \left. +\frac12 \sqrt{1-\beta(1-q_1)} \, 
	     \exp \left(\frac{\beta}{2}
	         (\eta^2-\frac{ z^2}{1-\beta(1-q_1)}) \right) 
	\left( \erf \, (\psi_+(0)) +\erf \, (\psi_-(0))
	\right)\right\}^k 
	\nonumber \\ 	    
   \label{eq:fetarsb}
\end{eqnarray}
with
\begin{equation}
 \psi_{\pm}(x) = \frac{[1- x(1-q_1)]\eta \pm  z}
                        {\sqrt{2(1-q_1)(1- x(1-q_1))}} \, ;
	\quad z= \sqrt{q_0}\, z_1 + \sqrt{q_1-q_0}\, z_2 \, .
	\label{eq:psi}	
\end{equation}
In the limit $\eta \rightarrow 0$ the expression reduces to the Hopfield
RSB1 free energy as calculated, e.g, in \cite{CAG,SK,BH}.

Again in the following we are only interested in the zero temperature
results. From the limit $\beta \rightarrow \infty$ of  expression 
(\ref{eq:fetarsb})  we can obtain the fixed-point equations for the 
relevant order parameters. Since the way to derive these formula is
standard and since their explicit expressions are algebraically
complicated we do not write them down. 

The zero temperature critical capacity $\alpha_c^{\RSB1}$ as a function
of $\eta$ is presented in Fig.~2. 
For $\eta=0$ we confirm the result $\alpha_c^{\RSB1}=0.13819$ found
in \cite{SK,BH}. For growing $\eta$ the results for $\alpha_c^{\RSB1}$
and $\alpha_c^{\RS}$ start deviating more. In view of the results on
the entropy (see Fig.~1) we expect that the difference keeps growing. 
Since the calculations are very tedious and since in the literature
one is mostly interested in values of $\eta$  smaller than $1$ 
\cite{ZLW1,ZLW2,In} we have plotted results up to $\eta=1$. For 
$\eta=1$, e.g.,  we find $0.16658$ for the RSB1 critical capacity versus 
$0.16384$ for the RS critical capacity. The results of \cite{ZLW1}
 overestimate this value.

\section{Concluding remarks}
\pin 
The replica method is applied to an existing neural network model with 
state-dependent couplings. Only those patterns having a correlation 
with the state of the system greater than a threshold $\eta$ contribute
to the couplings. 

The free energy is obtained and the fixed-point 
equations are studied at zero temperature.
It is shown that the fixed-point equations in the replica symmetric
approximation coincide with these found by the so-called heuristically
motivated statistical mean-field scheme developed in \cite{Ge,Pe},
provided one does not make the additional
assumption that the overlap with the non-condensed patterns is Gaussian.
This assumption, made in the literature, is totally unnecessary and is
even incompatible with the key assumption of the statistical method that 
the noise induced by the non-condensed patterns is Gaussian. 
 
The critical storage capacity at zero temperature is calculated as a 
function of the threshold $\eta$ and compared with the values obtained
in the literature on the basis of the statistical method with the extra
Gaussian assumption for the overlaps.  
Since a calculation of the entropy indicates that replica symmetry is
broken at zero temperature for all values of $\eta$
a first order replica symmetry breaking calculation has
been performed. The critical storage capacity increases versus the 
replica symmetric values but up to $\eta=1$ the increase is relatively
small.

\section*{Acknowledgments}
\pin
This work has been supported in part by the Research Fund of the
K.U.Leuven (Grant OT/94/9) and the Korea Science and Engineering
Foundation through the SRC program. The authors are indebted to
J.~Huyghebaert and G.~Jongen for constructive discussions.
One of us (D.B.) thanks the Belgian National Fund for Scientific
Research for financial support.

\end{document}